\documentclass[twocolumn, preprintnumbers, 
endnote,nofootinbib,prl,9pt]{revtex4}
\usepackage{graphicx}
\usepackage{subfigure}

\newcommand{\lsim}{\lesssim}
\newcommand{\gsim}{\gtrsim}

\newcommand{\eq}[1]{Eq.~(\ref{#1})}

\newcommand{\ord}[1]{\mathcal{O}{(#1)}}
\newcommand{\beq}{\begin{equation}}
\newcommand{\eeq}{\end{equation}}

\newcommand{\lD}{\Lambda_D}

\begin{document}

\pagestyle{plain}

\title{A 750 GeV Messenger of Dark Conformal Symmetry Breaking}

\author{Hooman Davoudiasl\footnote{email: hooman@bnl.gov}
}

\author{Cen Zhang\footnote{email: cenzhang@bnl.gov}
}

\affiliation{Department of Physics, Brookhaven National Laboratory,
Upton, NY 11973, USA}


\begin{abstract}

The tentative hints for a diphoton resonance at a mass of $\sim 750$~GeV from
the ATLAS and CMS experiments at the LHC may be interpreted as first contact
with a ``dark" sector with a spontaneously broken conformal symmetry.  The
implied TeV scale of the dark sector may be motivated by the interaction
strength required to accommodate a viable thermal relic dark matter (DM)
candidate.  We model
the conformal dynamics using a Randall-Sundrum type 5D geometry whose IR
boundary is identified with the dynamics of the composite dark sector, while
the Standard Model (SM) matter content resides on the UV boundary,
corresponding to ``elementary" fields.  We allow the gauge fields to reside in
the 5D bulk, which can be minimally chosen to be $SU(3)_c\times U(1)_Y$.  The
``dark" radion is identified as the putative 750 GeV resonance.  Heavy
vector-like fermions, often invoked to explain the diphoton excess, are not
explicitly present in our model and are not predicted to appear in the spectrum
of TeV scale states.  Our minimal setup favors scalar DM of $\ord{\text{TeV}}$
mass.  A generic expectation in this scenario, suggested by DM considerations,
is the appearance of vector bosons at $\sim$ few TeV, corresponding to the
gluon and hypercharge Kaluza-Klein (KK) modes that couple
to UV boundary states with strengths that are suppressed uniformly compared to
their SM values.  Our analysis suggests that these KK modes could be within the
reach of the LHC in the coming years.

\end{abstract}


\maketitle

Only time will tell whether the current intense interest in the hints for a
$\sim 750$~GeV diphoton resonance, implied by the ATLAS \cite{ATLAS750} and CMS
\cite{CMS750} data, is justified.  While the statistics are not reliable yet,
with currently $\sim 2\sigma$ and $\sim 1\sigma$ global significance from ATLAS
and CMS, respectively, the simplicity of the diphoton final state may argue for
some modest optimism, though any such attitude is not rigorously warranted.  In
any event, we adopt the more positive view of the potential hint and examine
what it may signify.

The scale of the putative resonance is tantalizingly close to the electroweak
scale and is aligned with expectations from ``naturalness" of the Higgs mass.
However, the dearth of evidence for the presence of massive electroweak
states, $W,Z,t,H$, in the signal does not make a connection
with electroweak symmetry breaking (EWSB) a natural inference.  Nonetheless, the TeV
scale can also be motivated from a very different perspective, namely the
observation that interactions governed by TeV scale particles can lead to
the correct order of magnitude abundance for the cosmic dark matter (DM); this is
often referred to as the ``WIMP" miracle.  While there are many different possibilities
that one can choose for the new physics, given the current hints, we will assume that the
$\sim 750$~GeV scale of the possible resonance is set by the dynamics of DM and
is not directly related to the physics of EWSB (see also Ref.~\cite{DM750}).

Conformal symmetry breaking provides an interesting arena for generation of new
mass scales and can often lead to the appearance of a ``light" scalar, the
dilaton, which could be its most accessible signal.  Given this motivation, we
will assume that the new resonance with mass $m_\phi \sim 750$~GeV is a dilaton of
a dark sector, which includes DM.  We
will use a Randall-Sundrum (RS) type 5D background \cite{Randall:1999ee} to
model the underlying physics, as a dual geometric description
\cite{RSHolography}, in whose context the radion $\phi_D$
\cite{Goldberger:1999uk,Csaki:2000zn,Csaki:2007ns} is the aforementioned
``dark" dilaton scalar.  For other recent work on the radion interpretation of
the 750 GeV diphoton excess, see also Ref.~\cite{radion750}.
For a variety of alternative approaches see, for example, Ref.~\cite{var}.

Given that current data suggests that the Standard Model (SM) is a weakly
interacting theory, made up of elementary degrees of freedom, we will confine
the matter content of the SM, including its Higgs sector, to the ultraviolet
boundary of the warped RS geometry.  We will allow the SM gauge sector to
propagate in the 5D bulk \cite{Bulkgauge}.  In an minimal setup, it suffices to
have only $SU(3)_c\times U(1)_Y$ in the bulk, which we will assume for now.
The composite sector, corresponding to fields that are localized near or at the
infrared (IR) boundary, are all assumed to be SM singlets, {\it i.e.} belong to
a dark sector, which could naturally include DM (for an earlier work with a similar
setup, see Ref.~\cite{Carmona:2015haa}).
Note that this arrangement assumes that the physics of EWSB, flavor, and
potentially other aspects of the SM are governed by the physics on the UV
boundary whose cutoff scale is much larger than $\sim$~TeV.  In particular, we
will not address the issue of the Higgs potential naturalness, which may be
associated with ``elementary" UV dynamics.

In the above setup, the radion $\phi_D$ will not have significant interactions
with the SM, except through ``volume suppressed" couplings to the SM bulk gauge
fields \cite{Csaki:2007ns}, from $SU(3)_c\times U(1)_Y$.  Note that since we
assume all SM matter to be confined to the UV boundary they do not affect the
radion couplings through loop effects.  In particular, our framework does not
include vector-like quarks in its spectrum of single particle states
\cite{hidgball}, which is different from many models that attempt to explain
the diphoton excess (see, for example, Refs.\cite{Harigaya:2015ezk,vfermion}).

Let the curvature scale of the 5D warped background be denoted by $k$ and the
fifth dimension have length $L$.  Note that identifying the UV scale as the
Planck mass $M_P\sim 10^{19}$~GeV would require $kL\sim 30$, since the IR scale
is given by $e^{-kL}\times \text{UV scale}$.  However, we may assume that the
UV boundary has a cutoff that is much lower, but well above the weak scale, so
that $kL\gg 1$.  In a minimal setup, the couplings of the radion $\phi_D$
to the SM gauge fields are then given by
\beq \frac{\phi_D}{4 kL\, \lD} (G^A_{\mu\nu} G^{A,\mu\nu} +
B_{\mu\nu}B^{\mu\nu})\,, \label{phiFF}
\eeq
where $\lD$ is the decay constant of the radion and provides the IR cutoff
scale; $B_{\mu\nu}$ and $G^A_{\mu\nu}$ are the hypercharge and color field
strengths, respectively.  We have ignored possible loop-induced UV localized
kinetic terms.  Since $B = \cos \theta_W \,\gamma - \sin \theta_W \,Z$, where
$\theta_W$ is the weak mixing angle, the above interaction yields the coupling
of $\phi_D$ to $\gamma\gamma$, $Z\gamma$, and $ZZ$, in the ratio
$\cos^2\theta_W$, $-\sin\theta_W \cos\theta_W$, and $\sin^2\theta_W$,
respectively.

We could also have $SU(2)_L$ in the bulk, by adding a $\phi_D
W^I_{\mu\nu}W^{I,\mu\nu}$ term.  With this term, $\phi_D$ would couple universally to
$\gamma\gamma$, $ZZ$ and $WW$, and in particular the $Z\gamma$ coupling
vanishes, indicating that the resonance will not show up in the $Z\gamma$ final
state.  Compared to the minimal setup, the branching ratio of $\phi_D\to
\gamma\gamma$ is increased by about $30\%$, so to produce the correct signal
strength other parameters of the model need to be modified at the $\sim10\%$
level.  Apart from this, there is no other significant difference in terms of
collider phenomenology, and so in the following we focus on the minimal setup
without a $\phi_D W^I_{\mu\nu}W^{I,\mu\nu}$ term.

The above interactions suffice to provide the production, through gluon fusion,
and decay, into photons, of the purported new resonance.  However, the ATLAS
data shows some mild preference for a resonance of width $\sim 45$~GeV, though
the evidence is not very strong.  If we take this preference seriously,
the interactions in \eq{phiFF} would not provide the needed width,
since we expect $\lD\gsim m_\phi$ and $kL\gg 1$.  However, there could in
principle be a large number of new massive modes that correspond to the
composite states, whose masses are generated by conformal symmetry breaking in
the IR (near the TeV scale).  These states are localized at the IR boundary and
hence would couple to the radion only suppressed by $1/\lD$.  If sufficiently
many of these states are lighter than $m_\phi/2$, they may provide widths of
$\ord{45}$~GeV.  Given that the evidence for the large width hypothesis is
quite modest, we will instead focus on the possibility of a narrow width for
$\phi_D$ and a minimal model content in our analysis.

The WIMP miracle motivates considering whether new dark composites can be good
DM candidates in our scenario.  Let us assume that the lightest such state with
mass $\lsim \lD$ is cosmologically stable due to some conserved charge or
parity.  We consider the cases of a Dirac fermion $X$ or a real scalar $\chi$, stabilized
with a suitable unbroken symmetry,
coupled to $\phi_D$ via \cite{Giudice:2000av,Csaki:2007ns,Davoudiasl:2010fb}
\beq
-\frac{\phi_D}{\lD}(m_X {\bar X} X - \partial_\mu \chi \partial ^\mu \chi
+ 2 m_\chi^2\, \chi^2)\,. \label{Xchi}
\eeq
As we will discuss later, one can achieve the correct relic abundance
for DM, through pair annihilation into a pair of $\phi_D$ final states.
The $\phi_D$
 final states then decay promptly into the SM, in our minimal
scenario.

While our model does not address the hierarchy problem, it still shares some of
the signals of the warped RS-type hierarchy and flavor models.  Namely, due to
the presence of the SM gauge fields in the 5D bulk, one expects that
Kaluza-Klein (KK) modes of these fields will appear at scales of order $\lD\sim$~TeV
(see also Ref.~\cite{Megias:2015ory}).
In particular, a gluon KK state of a few TeV mass may well be
within the reach of the LHC.  To see this, note that
the production of the gluon KK mode is very similar to the case of warped
models with light fermions localized near the UV boundary, to explain their
small masses (as in those models the Higgs is at the IR boundary).  However,
here, all quarks would couple to the KK gluon with the same strength and there
is no preference for top quarks.  Hence, the current bounds on RS KK gluons do
not directly apply.  However, with sufficient luminosity, one expects that KK
gluons of mass $\ord{\text{few TeV}}$ could be within the reach the LHC.  Other
gauge KK modes will also appear at the same mass scale, however their
production is suppressed by weak gauge couplings.  In our scenario, their
branching fraction into charged leptons is not suppressed compared to 
branching fractions into heavier
SM states and they may also be interesting targets for future searches at the
LHC.  Below, we will examine the possibility of looking for the hypercharge 
KK mode of our minimal model in the clean dilepton ($e^+e^-$ or $\mu^+\mu^-$) channel and find that it has 
discovery prospects comparable to that of the KK gluon.   
While generically present, the KK graviton -
which is $\sim 1.5$ heavier than gauge KK modes in the RS model \cite{Davoudiasl:1999jd} -
could potentially be outside the LHC reach.

{\bf Results:} As mentioned before, we have assumed a minimal model that is
consistent with a narrow scalar resonance at $\sim 750$~GeV.  To investigate
collider phenomenology, we use {\sc MadGraph5\_aMC@NLO} \cite{Alwall:2014hca},
with NN23LO1 PDF set \cite{Ball:2013hta}, and dynamical renormalization and
factorization scales set to one half of transverse mass summed
over all final states.  For the scalar resonance we always include an NLO
$K$-factor of $\sim1.2$ ($\sim1.4$) at 8 (13) TeV, obtained by using the Higgs
Characterisation model \cite{Artoisenet:2013puc}.  Our model is implemented
in the UFO format \cite{Degrande:2011ua} using the {\sc FeynRules} package
\cite{Alloul:2013bka}.  The width of $\phi_D$ and its branching ratios are computed with the {\sc MadWidth}
package \cite{Alwall:2014bza}.

We find that a signal strength of 5.9 fb, corresponding to an average between
ATLAS and CMS \cite{Buttazzo:2015txu}, with integrated luminosities of
3.2~fb$^{-1}$ and 2.6~fb$^{-1}$ respectively, can be obtained if $k L \lD
 \approx 40$~TeV.  Using the {\sc MadDM} package
\cite{Backovic:2013dpa}, we found that the $t$-channel $p$-wave annihilation,
$X\bar X\to \phi_D \phi_D$, will not readily yield an acceptable DM
abundance, $\Omega h^2 \approx 0.12$ \cite{Agashe:2014kda}, unless its mass is chosen very close to
$m_\phi/2$.  In this case, resonant annihilation through $s$-channel $\phi_D$
exchange can then be sufficiently strong, but the required $m_X$ is somewhat
tuned.  However, we find that a dark scalar $\chi$ from \eq{Xchi} can provide
the correct thermal relic density if it has a mass $m_\chi \approx 1$~TeV,
for $\lD\approx 5.5$~TeV.  The diphoton signal strength then implies $kL\approx 7$, and the UV
cutoff scale of the SM is hence given by $5.5\times e^7\sim 6000$~TeV, which
corresponds to a ``Little RS" geometry \cite{Davoudiasl:2008hx}.  In this
scenario, a KK gluon, $g_{KK}$, and a KK hypercharge gauge boson, $B_{KK}$,
both of mass $\lsim 5$~TeV, can be a reasonable expectation.

\begin{table}[htb]
	\begin{tabular}{l|cl}
		\hline
		\hline
		Parameters & $\Lambda_D$ & 5500 GeV \\
		& $kL$ & 7.23 \\
		& $m_{\phi}$ & 750 GeV \\
		& $M_\chi$ & 1040 GeV \\
		& $M^{g,B}_{KK}$ & 3000 GeV \\
		\hline
		Widths & $\Gamma_{\phi}$ & 0.012 GeV \\
		& $\Gamma_{g_{KK}}$ & 46.4 GeV\\
		& $\Gamma_{B_{KK}}$ & 12.7 GeV\\
		\hline
		Branching & Br($\phi_D\to\gamma\gamma$) & 6.54\% \\
		ratios & Br($\phi_D\to ZZ$) & 0.56\% \\
		& Br($\phi_D\to \gamma Z$) & 3.81\% \\
		& Br($\phi_D\to gg$) & 89.1\% \\
		& Br($g_{KK}\to q\bar q$) & 16.7\% \\
		& Br($B_{KK}\to l^+l^- $) & 10.0\% \\
		\hline
		Cross sections &
		$pp\to g_{KK}\to t\bar t$ & 103 fb\\
		(LHC 14 TeV) & $pp\to g_{KK} \to jj $ & 550 fb\\
		 & $pp\to B_{KK} \to e^+e^-,\mu^+\mu^- $ & 1.2 fb\\
		\hline
	\end{tabular}
	\caption{Benchmark point in the minimal model.  Here $q$ denotes a quark
		and $l$ is a charged lepton, of any flavor. \label{tab:b}}
\end{table}
A benchmark point is given in Table~\ref{tab:b} with more details.
This benchmark point could produce the correct signal strength and DM relic density.
We have checked that this point is consistent with 8 TeV resonance searches in
$\gamma\gamma$, $\gamma Z$, $ZZ$, and $jj$ final states
\cite{Aad:2015mna,Khachatryan:2015qba,Aad:2014fha,Aad:2015kna,Aad:2014aqa,CMS:2015neg}.    
Furthermore, the KK gluon resonance in $q\bar q$ and $t\bar t$ final states, 
and the KK hypercharge mode in the dilpeton final state 
are consistent with 8 TeV (as well as 13 TeV for dileptons) searches
\cite{Aad:2014aqa,Chatrchyan:2013lca,Aad:2014cka,dilepton13}.
The coupling of the KK gauge fields to UV-localized fields is well-estimated by
$1.2 g/\sqrt{k L}$ in our model \cite{Agashe:2007ki}, where $g$ is the relevant
coupling constant.

We find that at the 14 TeV LHC, the 3~TeV KK gluon can be produced in the
$t\bar t$ final state, with a cross section of $\sim 100$ fb, well above the
reach for $t\bar t$ resonance search at 14 TeV, which is $\sim$ 10-20~fb
in the all-hadronic channel with $300$ fb$^{-1}$ of integrated luminosity
\cite{Agashe:2013kyb}.  Assuming a $S/\sqrt{B}$ scaling, where $S$ denotes signal and 
$B$ is background, we estimate that the benchmark 3~TeV KK gluon can be discovered with 
$\ord{10}$~fb$^{-1}$ of integrated luminosity.  
Alternatively, the KK gluon can decay into two jets.  With
our benchmark coupling, discovery potential for a color octet vector in the di-jet
final state can reach $\sim 4$ TeV with $300$ fb$^{-1}$ of integrated luminosity
\cite{Yu:2013wta}.  Thus within this scenario, based on DM considerations, KK
gauge fields can be expected to be within the reach of the 14 TeV LHC with
$\mathcal{O}(10)$ or more fb$^{-1}$ of integrated luminosity.

As for a 3 TeV KK hypercharge state, we find that the cross section for 
$pp\to B_{KK}\to \text{dilepton}$ at the 14 TeV LHC is about 1~fb, with negligible
background \cite{dilepton13}.  
Hence, for a handful of events, assuming an efficiency of $\sim 50\%$, we would need 
$\ord{10}$~fb$^{-1}$.  We then see that the prospect for discovery of the KK gluon in 
the $t\bar t$ and the KK hypercharge in the dilepton channels are comparable.

We note that our framework can trivially include bulk singlet fermions corresponding to
right-handed neutrinos, localized near the IR boundary, to achieve natural Dirac masses for
neutrinos \cite{Grossman:1999ra}.  Alternatively, we may include UV-boundary heavy Majorana neutrinos  with masses
near the cutoff scale, $M_N\lsim 6\times 10^3$~TeV, to yield seesaw masses for light
neutrinos, assuming Yukawa couplings $\sim 10^{-4}$, similar to those of light SM fermions.

In conclusion, we have proposed that the $\sim 750$~GeV diphoton excess,
reported by ATLAS and CMS, can be due to a dilaton scalar, associated with dark
conformal symmetry breaking.  The dynamics of the conformal sector can also
provide a DM candidate.  Using a dual 5D RS-type geometric description, the
requisite couplings of the ``dark'' radion, identified as the diphoton resonance, can
be achieved by assuming that the gauge sector of the SM propagates in the 5D
bulk.  We assume that the rest of the SM corresponds to elementary fields that
are localized at the UV boundary.  We find that an IR-localized scalar of $\sim
1$~TeV mass can be a suitable DM candidate if the scale that sets the coupling
of the radion is about 5 TeV.  In this setup, we may then expect that the KK
gauge modes could be within the reach of the LHC Run~II with
$\ord{10}$~fb$^{-1}$ or more of integrated luminosity.

\acknowledgments

H.D. thanks K. Agashe for discussions.  This work is supported in part by the
United States Department of Energy
under Grant Contracts DE-SC0012704.

\end{document}